\documentclass[twoside]{dis07}
\usepackage[latin1]{inputenc}
\usepackage[dvips]{graphicx,epsfig,color}
\usepackage{wrapfig,rotating}
\usepackage{cite}
\usepackage{amssymb,amsmath,array}

\newcommand\GeV{\,\mbox{${\rm GeV}$}\,}
\newcommand\MeV{\,\mbox{${\rm MeV}$}\,}

\begin{document}
\title{{\normalsize\sl DESY 07/083     \hfill {\tt arXiv:yymm.nnnn}
\\ SFB/CPP-07-28 \hfill {   } }\\
\boldmath{$\Lambda_{\rm QCD}$} and \boldmath{$\alpha_s(M_Z^2)$} from DIS 
Structure Functions}

\author{Johannes Bl\"umlein$^1$ 
%
\thanks{This paper was supported in part by SFB-TR-9: Computergest\"utze 
Theoretische Teilchenphysik.}
%
\vspace{.3cm}\\
%
Deutsches Elektronen-Synchrotron, DESY,
Platanenallee 6, D-15738 Zeuthen, Germany
}

\maketitle

\begin{abstract}
A brief summary is given on recent determinations of $\Lambda_{\rm QCD}$ 
and $\alpha_s(M_Z^2)$ from deeply inelastic structure functions.
\end{abstract}


\noindent
Various QCD analyzes of the world unpolarized and polarized 
deep-inelastic data on charged 
lepton--nucleon  and neutrino--nucleus scattering were performed in order 
to measure the QCD scale $\Lambda_{\rm QCD}$, resp. $\alpha_s(M_Z^2)$, 
from the scaling violations of the nucleon structure functions. In this note 
we give a brief overview on the status of these analyzes.\footnote{For a 
recent survey on the status of deep-inelastic scattering see \cite{SURV}.} 
Most of the analyzes
performed in the past were of next-to-leading order, see Table 1. Here the 
values of $\alpha_s(M_Z^2)$ range between 0.1171--0.1148 mainly with the 
exception of the old, very low BCDMS value \cite{BCDMS} and 
Ref.~\cite{GRS} obtaining $\alpha_s(M_Z^2) = 0.112$.
The typical 
theory error is estimated by varying the renormalization and 
factorization scales between $Q^2/4$ and $4 \cdot Q^2$ amounts to 
$\sim$5\% for 
$\alpha_s(M_Z^2)$, a theoretical uncertainty too large to cope with the 
experimental uncertainty of $O(1..2\%)$ after the completion of the HERA 
programme. 
The analyzes of polarized nucleon data still yield rather large 
errors \cite{BB,ALT} due to the present accuracy reached for 
polarization asymmetries. Moreover, these analyzes include data down to 
$Q^2 \sim 1 \GeV^2$, which is not unproblematic w.r.t. higher twist terms, 
the scaling violations of which are yet unknown.
The unpolarized analyzes at the present  level of accuracy require 
rigorous cuts for
potential higher twist effects, which can be achieved demanding $W^2 > 12 
... 15 \GeV^2$. Furthermore, we will consider only proton and deuteron data,
to avoid potential interference with nuclear effects.

With the advent of the 3-loop anomalous dimensions \cite{THRL} in the 
unpolarized case one may 
extend the analysis to next-to-next-to-leading order, where the remaining 
theory error is of $O(1\%)$ or less, see below. To cope with the present 
experimental errors 3--loop analyzes are mandatory. 
A theoretically consistent 
analysis can be performed at least in the 
non-singlet case, where the heavy flavor effects known to $O(\alpha_s^2)$,
are negligibly small. 3--loop valence analyzes were performed in 
\cite{BBG,GRS}. One even may extend the non-singlet analysis to the 
4--loop level \cite{BBG}. A closer numerical study of the potential effect 
of the 
yet missing 4--loop anomalous dimension, performing a comparison with the
recently calculated second moment in \cite{BC} shows that the overwhelming 
effect at 4--loops is due to the 3--loop Wilson coefficient. 
To see the convergence of the perturbative expansion we list the values for 
$\alpha_s(M_Z^2)$ obtained in the NLO, N$^2$LO, and N$^3$LO analyzes~:
\begin{equation}
\alpha_s(M_Z^2) = 0.1148 \rightarrow 0.1134 \rightarrow 0.1142 \pm 0.0021.
\end{equation}
The change from the N$^2$LO to the N$^3$LO value is found deeply inside 
the current experimental error. The  N$^3$LO value corresponds to 
\begin{equation}
\Lambda_{\rm QCD}^{\rm \overline{MS}, 
N_f =4} = 234 \pm 26 \MeV.
\end{equation}
A drawback of the valence analysis are small, remaining contributions of sea-quark 
densities in the region $x > 0.4$, the effect of which can finally only be studied
in combined singlet/ non-singlet analyzes.

In the singlet case the 3--loop heavy flavor corrections are yet missing.
Still analyzes may be performed to determine $\Lambda_{\rm QCD}$ under 
an assumption for these terms. The results are summarized in Table~1.
Compared to the respective NLO analyzes, the values of $\alpha_s(M_Z^2)$ turn out
to be lower by 1--2\% in case comparable values are available. Three independent analyses 
using different codes and methods to solve the evolution equations agree 
\cite{GRS,BBG,A06} at the $1\sigma$ level and better. 
These analyzes were performed 
using the world structure function data for deep-inelastic charged lepton proton and deuteron 
scattering.  
The analysis in \cite{A06} is a combined singlet and non--singlet analysis and
fully confirms the value of $\alpha_s(M_Z^2)$ obtained in the non--singlet
analysis Ref.~\cite{BBG}, showing that the remaining uncertainties there do not affect the
value of $\Lambda_{\rm QCD}$. Alternatively to the standard $\overline{\rm MS}$-analysis
one may perform factorization scheme-invariant analyzes \cite{SI}, based on observables only.
This method is free of shape--assumptions, in particular for the gluon density.
A slightly higher value of $\alpha_s(M_Z^2)$ was found in an earlier analysis \cite{SY} using 
the method of Bernstein polynomials. A recent analysis \cite{THR}, including also jet data from 
colliders, reports a much higher value of $\alpha_s(M_Z^2)$.

$\Lambda_{\rm QCD}^{\overline{\rm MS}}$ was measured also in two recent 
lattice simulations based on two active flavors ($N_f =2$). These 
investigations paid 
special attention to non-perturbative renormalization and kept the 
systematic errors as small as possible. 
\begin{eqnarray}
\Lambda_{N_f=2}^{\rm latt} = 245 \pm 16 \pm 16 \MeV~~[17],~~~~~~ 
\Lambda_{N_f=2}^{\rm latt} = 261 \pm 17 \pm 26 \MeV~~[18] 
\end{eqnarray}
\begin{center}
\begin{tabular}{|l|lll|c|}
\hline
\hline
 {\bf NLO} & $\alpha_s(M_Z^2)$ & expt & theory & Ref. \\
\hline
\hline
 CTEQ6    & 0.1165 & $\pm$0.0065 &             & \cite{CTEQ6} \\
 A02      & 0.1171 & $\pm$0.0015 & $\pm$0.0033 & \cite{A02} \\
 ZEUS     & 0.1166 & $\pm$0.0049 &             & \cite{ZEUS} \\
 H1       & 0.1150 & $\pm$0.0017 & $\pm$0.0050 & \cite{H1} \\
 BCDMS    & 0.110  & $\pm$0.006  &             & \cite{BCDMS} \\
 GRS      & 0.112  &             &             & \cite{GRS} \\
 BBG      & 0.1148 & $\pm$0.0019 &             & \cite{BBG} \\
\hline
 BB (pol) & 0.113  & $\pm$0.004  & \hspace*{-0.25cm}
\tiny{$\begin{array}{c} +0.009 \\ -0.006 \end{array}$} & \cite{BB} \\
\hline \hline
 {\bf N$^2$LO}   & $\alpha_s(M_Z^2)$           & expt & theory& Ref. \\
\hline
\hline
 A02m         & 0.1141 & $\pm$0.0014         & $\pm$0.0009 & \cite{A06}  \\
 SY01(ep)     & 0.1166 & $\pm$0.0013         &             & \cite{SY}  \\
 MSTW         & 0.1191 &  $\pm$0.002         & $\pm$0.003  & \cite{THR}  \\
 GRS          & 0.111  &                     &             & \cite{GRS}  \\
 A06          & 0.1128 & $+0.0015$           &             & \cite{A06}  \\
 BBG          & 0.1134 & $+0.0019/-0.0021$   &             & \cite{BBG}  \\
\hline \hline
{\bf N$^3$LO} &        &                     &             &      \\
 BBG          & 0.1142 & $\pm$ 0.0021        &             & \cite{BBG}  \\
\hline
\hline
\end{tabular}
\end{center}

\vspace{2mm} \noindent
\begin{center}
{\sf Table~1:~ Summary of $\alpha_s(M_Z^2)$ values determined from deep-inelastic scattering.
}
\end{center}
\vspace{2mm} \noindent

A direct comparison with the case $N_f = 4$ in the 
above data analyzes is not yet possible. However, the 
difference between the earlier $N_f = 0$ and the present result in 
$\Lambda_{\rm QCD}$ amounts to $O(10 \MeV)$
only. We have to wait and see what is obtained for $N_f = 4$ in coming 
analyzes. 

More global analyzes were performed using also semi-inclusive $ep$--
and $pp$--data from jet measurement, mostly aiming on a global 
determination of the quark and gluon densities. As shown in \cite{SB,CTEQ,AL_DIS,MCS} 
the $\alpha_s(M_Z^2)$ values obtained in 
analyzing the jet data and other data sets beyond those of the 
structure functions differ significantly in their $\chi^2$--profiles and
fitted value for the strong coupling constant pointing to systematic 
differences. The jet data prefer a higher value of $\alpha_s(M_Z^2)$ than 
the inclusive DIS data. This effect deserves further detailed studies before
one is allowed to combine these data sets for a precision determination of 
$\Lambda_{\rm QCD}$.
 
%
%
%
%
\begin{footnotesize} 

\end{footnotesize}
\end{document}